\documentclass[prb,preprint]{revtex4-2}

\usepackage{amsmath}  
\usepackage{amsfonts} 
\usepackage{graphicx} 

\begin{document}

\title{Comment on ``Specific heat of an ideal Bose gas above the Bose
  condensation temperature,'' [Am.~J.~Phys.~\textbf{72}(9), 1193--1194
    (2004)]}

\author{Frank Wang} \email{fwang@lagcc.cuny.edu}
\affiliation{Department of Mathematics, LaGuardia Community College of
  the City University of New York, Long Island City, NY 11101}

\date{\today}

\maketitle 

The graph of the specific heat of an ideal Bose gas as a function of
temperature exhibits a characteristic $\Lambda$-shaped behavior near
the critical temperature of Bose-Einstein condensation.  While there
is an explicit formula for the specific heat below the critical
temperature, I failed to find that for above the critical temperature
in standard textbooks.  In 2004, I presented a derivation for an
approximate formula in the \textit{American Journal of
  Physics}.\cite{fw2004} In 2025, Princeton University Press published
\textit{The Essential Einstein},\cite{ee2025} a collection of Albert
Einstein's most important works including an English translation of
Einstein's groundbreaking 1925 paper,\cite{einstein1925} and from it I
learned that Einstein himself set a procedure for the formula I was
searching for.  In this Comment I will guide readers to execute the
calculations that Einstein outlined, correct some numerical errors in
the paper, and compare his formula with mine.  The history of the
acceptance of Einstein's theory will be summarized.

\begin{figure}
  \includegraphics[scale=0.6,angle=0]{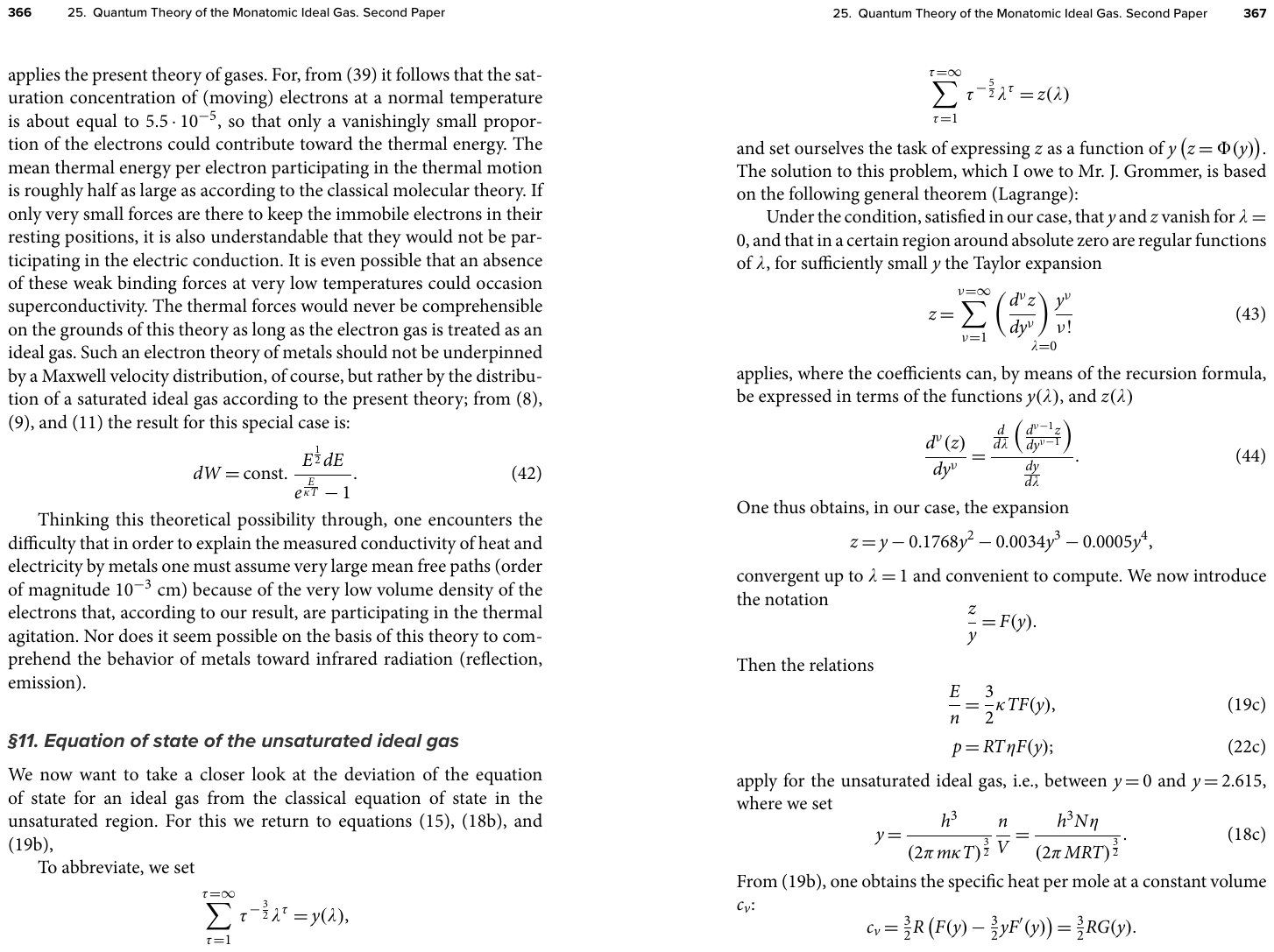}
  \caption{\label{fig:ee367} From \textit{The Essential
    Einstein}\cite{ee2025}.  Courtesy of Tilman Sauer.}
\end{figure}

In 1924, Einstein published ``Quantum Theory of the Monoatomic Ideal
Gas'' (Ref.~\cite{ee2025}, Chapter 24) which laid the foundation for
quantum statistical mechanics.  Einstein's interest was triggered by
the request to translate (to German) and publish a paper in
\textit{Zeitschrift f{\"u}r Physik}, previously rejected by the
\textit{Philosophical Magazine}, by the then unknown Indian physicist
Satyendranath Bose.\cite{bose1924} Bose used an innovative way to
count photons and obtained the Planck distribution;\cite{bose1924}
Einstein generalized Bose's counting method for identical particles,
and established what is now known as the Bose-Einstein distribution.
In 1925, Einstein wrote a second paper with the same title
(Ref.~\cite{ee2025}, Chapter 25) and predicted the phenomenon of ``Das
ges{\"a}ttigte ideale Gas'' (the saturated ideal gas) which is now
referred to as the phenomenon of Bose-Einstein condensation.  He ended
the paper with the equation of state of the unsaturated ideal gas (see
Figure~\ref{fig:ee367}) which is the focus of this note.

Einstein used $\overline{E}$ for the internal energy and the lowercase
$n$ for the total number of bosons, and he wrote (Eq.~(22) in
Ref.~\cite{ee2025} on page 353)
\begin{equation}\label{eq:eq1}
  \frac{\overline{E}}{n} = \frac{3}{2} \kappa T
  \frac{\sum_{\tau=1}^{\infty} \tau^{-5/2} \lambda^{\tau}}
       {\sum_{\tau=1}^{\infty} \tau^{-3/2} \lambda^{\tau}} ,
\end{equation}
where he used $\kappa$ for the Boltzmann constant, $T$ for
temperature, and $\lambda$ is the fugacity $e^{\mu/\kappa T}$ (where
$\mu$ is the chemical potential), which is a real number between 0 and
1.  One can read Einstein's original papers or Ref.~\cite{fw2004} for
this result; his internal energy is equivalent to Eq.~(17) in
Ref.~\cite{fw2004}; see the Appendix.  Einstein denoted the
denominator in Eq.~(\ref{eq:eq1}) by $y(\lambda)$ and the numerator by
$z(\lambda)$.  His $y$ in his Eq.~(18c) shown in
Figure~\ref{fig:ee367} is essentially a temperature parameter (with
fixed number of particles $n$ and volume $V$) and can be written more
compactly as
\begin{equation}\label{eq:ycomp}
  y=\zeta(3/2) \left( \frac{T_{c}}{T} \right)^{3/2} ,
\end{equation}
where $\zeta(3/2)$ is the Riemann zeta function $\zeta(s) =
\sum_{n=1}^{\infty} n^{-s}$ evaluated at $3/2$ and $T_{c}$ is the
critical temperature for Bose-Einstein condensation; see
Ref.~\cite{fw2004} Eq.~(11).  The value of $\zeta(3/2)$ is $2.612375
\ldots$ but Einstein used a crude value of $2.615$ for it throughout
his paper.  (The value of $\zeta(5/2)$ is $1.341487 \ldots$ but
Einstein used $1.348$ in his Eqs.~(40) and (41).  An examination of
Einstein's handwritten manuscript\cite{einstein1924h} revealed
mistakes introduced by the typesetter: in Eq.~(40), a factor of $3/2$
for the internal energy is missing, and in Eq.~(18c), the Avogadro
constant $N$ should be raised to the 4th power.)

Einstein sought to express $z$ as a function of $y$, and he credited
his assistant Jakob Grommer for the solution using the Lagrange
inversion theorem.\cite{hurwitz1922} For sufficiently small $y$ one
can calculate a Taylor expansion of $z$ in $y$, and the coefficients
can be obtained from repeated differentiation.  In the present case
the differentiation $d^{\nu} z/dy^{\nu}$ involves parametric equations
$z(\lambda)$ and $y(\lambda)$, but the procedure is a standard topic
in undergraduate calculus.  Einstein's Eqs.~(43) and (44),
$$
z = \sum_{\nu=1}^{\infty} \left(
\frac{d^{\nu}z}{dy^{\nu}} \right)_{\lambda=0} \frac{y^{\nu}}{\nu !} ,
\quad
\frac{d^{\nu} z}{d y^{\nu}} =
\frac{\frac{d}{d\lambda}\left( \frac{d^{\nu-1} z}{d y^{\nu-1}} \right)}
     {\frac{dy}{d \lambda}}
$$
can be implemented using a computer algebra system such as Maple or
Mathematica, which excels in this kind of repetitive and tedious
process.
\begin{equation}\label{eq:zofleq}
  z = y - \frac{1}{2!} \frac{1}{2 \sqrt{2}} y^{2} -
  \frac{1}{3!} \left( \frac{4}{3 \sqrt{3}} - \frac{3}{4} \right) y^{3}
  - \frac{1}{4!} \left(\frac{9}{4} + \frac{15 \sqrt{2}}{8} - 2 \sqrt{6}
  \right) y^{4} + \ldots ,
\end{equation}
or
\begin{equation*}
  z = y -0.1768 \, y^{2} - 0.0033 \, y^{3} - \, 0.00011 y^{4} + \ldots .
  \tag{\ref{eq:zofleq}$^{\prime}$}
\end{equation*}
Einstein's quartic term in his expansion
$$z = y - 0.1768 \, y^{2} - 0.0034 \, y^{3} - 0.0005 \, y^{4}
$$
shown in Figure~\ref{fig:ee367} is inaccurate; it is conceivable that
he or his assistant committed an error while performing the laborious
manual calculation.  (Einstein's coefficient for the quadratic term,
0.1768, is correct here, but he apparently transcribed it erroneously
to his Eq.~(22d) by writing 0.186.)

\begin{figure}
  \includegraphics[scale=0.6]{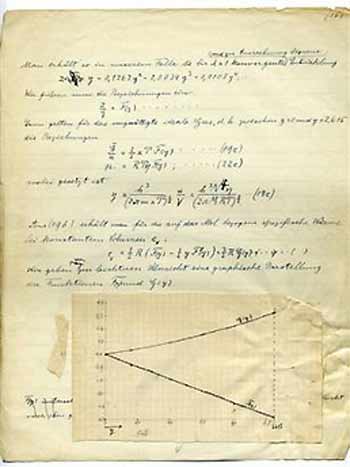}
   \hspace{16pt}
  \includegraphics[scale=0.6]{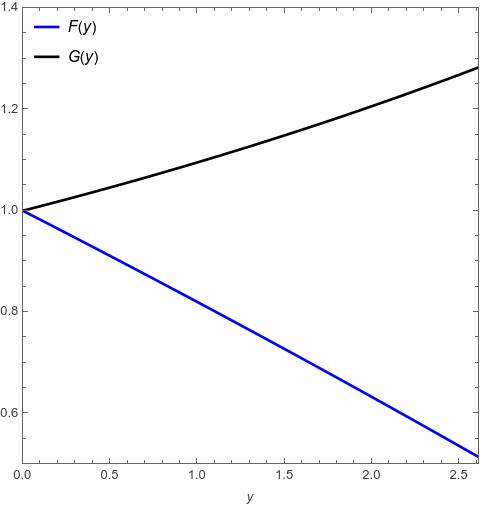}
  \caption{\label{fig:EinsteinFG} Einstein's handwritten manuscript
    has a graph of the function $F(y)$ and $G(y)$ for an easier
    overview;\cite{einstein1924h} the same graph with corrected values
    is recreated using computer software.}
\end{figure}

He introduced the notation $F(y)=z/y$, and with it the internal energy
per particle is $3/2 \, \kappa T F(y)$, see his Eq.~(19c). (A bar over
$E$ is missing in the English translation.)  The specific heat is the
derivative of the internal energy with respect to temperature, and it
involves the following derivative rules.
$$
  \frac{d}{dT} \left[ T F(y) \right] = F(y) + T \frac{d F(y)}{dT} =
  F(y) + T \frac{d F(y)}{dy} \frac{dy}{dT} .
$$
Because $y$ is proportional to $T^{-3/2}$, $T \, dy/dT = -3/2 \, y$.
Einstein defined a function $G(y)$, and based on the correct
$z$ in Eq.~(\ref{eq:zofleq}), it should be
\begin{equation}
  G(y) = F(y) - \frac{3}{2} y F^{\prime}(y)
  = 1 + 0.00884 y + 0.0066 y^{2} + 0.00039 y^{3},
\end{equation}
which he did not explicitly write down.  Nevertheless, he created a
graph, the only one in this paper, of the functions $F(y)$ and $G(y)$;
see Figure~\ref{fig:EinsteinFG}.  (In Einstein's notation, the
classical ideal gas law would be $p=RT \eta$, where $p$ is the
pressure, $R$ is the ideal gas constant, and $\eta = n/(NV)$ is mole
per volume.  For bosons, the law is expressed as the virial expansion
$p=RT \eta F(y)$; see Einstein's Eq.~(22c) in Figure~\ref{fig:ee367}
or Figure~\ref{fig:EinsteinFG}.)

Recall from Eq.~(\ref{eq:ycomp}) that $y$ is a temperature parameter,
thus the specific heat per mole above $T_{c}$ as a function of $T$
based on Einstein's approach (see $c_{v}$ after his Eq.~(18c) in
Figure~\ref{fig:ee367}) after correcting the numerical error in his
$z$ is
\begin{equation}\label{eq:einsteincv1925}
  c_{v} = \frac{3}{2} R \, G(y) = 
  \frac{3}{2} R \left[ 1 + 0.231 \left(\frac{T_{c}}{T}\right)^{3/2}
    + 0.045 \left(\frac{T_{c}}{T}\right)^{3} + 0.0069
    \left(\frac{T_{c}}{T}\right)^{9/2} \right] .
\end{equation}
For comparison, the specific heat above $T_{c}$ in Ref.~\cite{fw2004}
is reproduced here.
\begin{equation}\label{eq:wangcv}
  c_{v} = R \left\{ \left[ \frac{9}{2} \frac{\zeta(5/2)}{\zeta(3/2)}
    -\frac{3}{8} \frac{\zeta(3/2)^{2}}{\pi} \right] +
  \left[\frac{9}{4} \frac{\zeta(5/2)}{\zeta(3/2)} -
    \frac{3}{8} \frac{\zeta(3/2)^{2}}{\pi} \right]
    \left( \frac{T_{c}}{T} \right)^{3/2} +
  \left[\frac{3}{4} \frac{\zeta(3/2)^{2}}{\pi}
    -\frac{3 \zeta(5/2)}{\zeta(3/2)} \right]
  \left( \frac{T_{c}}{T} \right)^{3}
  \right\} ,
\end{equation}
or
\begin{equation}
  c_{v} = \frac{3}{2} R \left[ 0.997 + 0.227 \left( \frac{T_{c}}{T}
    \right)^{3/2} + 0.057 \left( \frac{T_{c}}{T} \right)^{3}
    \right].   \tag{\ref{eq:wangcv}$^{\prime}$}
\end{equation}

\begin{figure}
\includegraphics[scale=0.6]{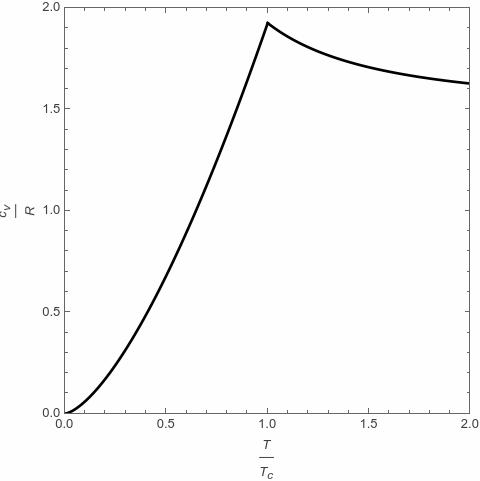}
\caption{\label{fig:ewcv} Specific heat of an ideal Bose-Einstein gas;
  Eq.~(\ref{eq:below}) for $0 \le T/T_{c} \le 1$ and
  Eq.~(\ref{eq:wangcv}) for $T/T_{c} \ge 1$.}
\end{figure}

The specific heat below $T_{c}$ is well known; see Eq.~(1) in
Ref.~\cite{fw2004}.
\begin{equation}\label{eq:below}
  c_{v} = \frac{15}{4} R \left( \frac{T}{T_{c}} \right)^{3/2}
  \frac{\zeta(5/2)}{\zeta(3/2)} = 1.926 \, R
  \left( \frac{T}{T_{c}} \right)^{3/2} .
\end{equation}
Using this result, the specific heat of an ideal Bose-Einstein gas as
a function of temperature is shown in Figure~\ref{fig:ewcv}.  In this
plot, it is visually indistinguishable whether one uses Einstein's
Eq.~(\ref{eq:einsteincv1925}) or my Eq.~(\ref{eq:wangcv}) for the
specific heat above $T_{c}$, but the two formulas are fundamentally
different.  Einstein made the expansion around $y=0$, which is the
limit $T \rightarrow \infty$, while I made the expansion around
$T=T_{c}$.  Einstein's specific heat tends to $3/2 \, R$ as $T$ tends
to infinity, while my specific heat and the discontinuity of the first
derivatives with respect to temperature $(\partial c_{v}/\partial T)$
at $T_{c}$ are \textit{exact} in terms of $\zeta(3/2)$, $\zeta(5/2)$,
and $\pi$.  Coincidentally, Einstein's high-temperature formula works
fine near the critical temperature, although one cannot ascertain
where or not the specific heat is continuous at $T_{c}$, and my
critical-temperature formula is satisfactory at high temperature.

After Einstein's 1925 publication, ``the degeneracy of the
Bose-Einstein gas has rather got the reputation of having only a
purely imaginary existence''\cite{london1938a} and scientists
overlooked this work for years.  In 1938, P.~Kapitza and independently
J.~F.~Allen and D.~Misener published results of the superfluidity of
liquid helium.  The same year Fritz London wrote a letter to
\textit{Nature}\cite{london1938a} with a plot of the specific heat
similar to Figure~\ref{fig:ewcv}; presumably he used
Eq.~(\ref{eq:london1938a}) below.  The resemblance between the
$\Lambda$ shape and the measured specific heat convinced most
physicists about a direct link between Einstein's theoretical work and
experimental observables.  In a 1939 essay written for
\textit{Harper's Magazine} titled ``The Usefulness of Useless
Knowledge''\cite{flexner1939} by Abraham Flexner, the then director of
the Institute for Advanced Study in Princeton, he highlighted
Einstein's quantum statistical research as an example that unexpected
applications might emerge from seemingly ``useless'' studies.  He
quoted the following from \textit{Science}: ``The stature of Professor
Albert Einstein's genius reached new heights when it was disclosed
that the learned mathematical physicist developed mathematics fifteen
years ago which are now helping to solve the mysteries of the amazing
fluidity of helium near the absolute zero of temperature scale.''

In London's letter to \textit{Nature}, he wrote
\begin{equation}\label{eq:london1938a}
  c_{v} = \frac{3}{2} R \left[ 1 + 0.231
    \left(\frac{T_{c}}{T}\right)^{3/2} + 0.046
    \left(\frac{T_{c}}{T}\right)^{3} + \ldots \right]
\end{equation}
for $T \ge T_{c}$.  Comparing this with Eq.~(\ref{eq:einsteincv1925}),
London seemed to have noticed the numerical error in Einstein's 1925
paper and dropped the last term.  London published a paper in
\textit{Physical Review} titled ``On the Bose-Einstein
Condensation,''\cite{london1938b} and he stated ``for $T \ge T_{c}$,
Einstein has previously given the semi-convergent expansion.''  The
specific heat in London's paper is
\begin{equation}
  c_{v} = \frac{3}{2} R \left[ 1 +
    0.231 \left(\frac{T_{c}}{T}\right)^{3/2}
    + 0.045 \left(\frac{T_{c}}{T}\right)^{3} + 0.040
    \left(\frac{T_{c}}{T}\right)^{9/2} + \ldots \right] .
\end{equation}
The coefficient of the $T^{-9/2}$ term was incorrect.  Finally, in
London's monograph \textit{Superfluids}\cite{london1954} published in
1954, he wrote a formula almost identical to
Eq.~(\ref{eq:einsteincv1925}), except he rounded $0.0069$ to $0.007$
in the $T^{-9/2}$ term.  Like Einstein, London made the expansion
around the limit $T \rightarrow \infty$, as I pointed out in
Ref.~\cite{fw2004}.

In his book \textit{Einstein and the Quantum},\cite{stone} Douglas
Stone wrote, ``The generosity of the `Bose-Einstein' designation is
not widely appreciated, as few physicists realize that Bose said not a
word about the quantum ideal gas in his seminal paper.  The paper that
does predict quantum condensation belongs to Einstein alone.''  Bose's
1924 paper has been translated back to English in this
\textit{Journal},\cite{bose1924} and in the quote above Stone is
referring to Einstein's 1925 paper.\cite{einstein1925} The readers can
assess Stone's statement based on the original publications.  The term
``Bose-Einstein condensation'' first appeared in the title of London's
1938 \textit{Physical Review} paper,\cite{london1938b} and most papers
and books have since then adopted this usage.  When I submitted my
manuscript to the \textit{American Journal of Physics} in 2004, I
followed London's word choice of ``Bose-Einstein condensation,'' but
the editor abbreviated my title to ``Bose condensation'' in the
published version.

\begin{acknowledgments}
  I thank Ian Lennon Anton for bringing \textit{The Essential
    Einstein} to my attention.
\end{acknowledgments}

\appendix*
\section{Properties of the Bose-Einstein Integral Function}
In the main text I stated Einstein's expression of the internal
energy, Eq.~(\ref{eq:eq1}), is equivalent to Ref.~\cite{fw2004}
Eq.~(17), which reads
\begin{equation}
  U = \frac{3}{2} N k T \frac{g_{5/2}(\alpha)}{g_{3/2}(\alpha)} ,
\end{equation}
where $U$ is the internal energy and the Bose-Einstein integral
function is defined as
\begin{equation}
  g_{\sigma} (z) = \frac{1}{\Gamma(\sigma)} \int_{0}^{\infty}
  \frac{x^{\sigma-1}}{z^{-1} e^{x} - 1} \, dx,
\end{equation}
where $\Gamma(\sigma)$ is the Gamma function.  Do not confuse the
fugacity $z$ used here and in Ref.~\cite{fw2004} with Einstein's
definition of $z$ in the main text.  For small $z$, which represents
high temperature and low density, we can expand the integrand as a
power series then perform the improper integral term by term.
\begin{equation}
  g_{\sigma} (z) = \frac{1}{\Gamma(\sigma)} \int_{0}^{\infty}
  x^{\sigma-1} \sum_{l=1}^{\infty} (z e^{-x})^{l} \, dx =
  z + \frac{z^{2}}{2^{\sigma}} + \frac{z^{3}}{3^{\sigma}} +
  \ldots = \sum_{l=1}^{\infty} l^{-\sigma} z^{l} .
\end{equation}
This is the connection between the Bose-Einstein integral and the
summation that Einstein used for the internal energy in
Eq.~(\ref{eq:eq1}) in the main text.  But if one is interested in low
temperature regime, this series is not appropriate, because $z \equiv
e^{\beta \mu}$ is actually an exponential function.  This is a
Dirichlet series, named after the mathematician Peter Gustav Dirichlet
(1805--1859), who married Rebecka Mendelssohn, the younger sister of
the composers Fanny and Felix.  In Ref.~\cite{fw2004} a different
series has been developed to approximate the specific heat near the
critical temperature of Bose-Einstein condensation.

\end{document}